# Imaging the antiparallel magnetic alignment of adjacent Fe and MnAs thin films.

R. Breitwieser[1,2], M. Marangolo[1], J. Lüning[2,3], N. Jaouen[3], L. Joly[4], M. Eddrief[1], V. H. Etgens[1] and M. Sacchi[2,3]

1- INSP, UPMC Paris 06, CNRS UMR 7588, 140 rue de Lourmel, 75015 Paris, France
2- Laboratoire de Chimie Physique – Matière et Rayonnement, UPMC– Paris 06, CNRS UMR 7614, 11 Rue Pierre et Marie Curie, 75005  Paris, France
3- Synchrotron SOLEIL, B.P. 48, 91192 Gif–sur–Yvette, France
4- Swiss Light Source, Paul Scherrer Institut, 5232 Villigen PSI, Switzerland.

Abstract: The magnetic coupling between iron and α–MnAs in the epitaxial system Fe/MnAs/GaAs(001) has been studied at the sub-micron scale, using element selective x-ray photoemission electron microscopy. At room temperature, MnAs layers display ridges and grooves, alternating α (magnetic) and β (non-magnetic) phases. The self-organised microstructure of MnAs and the stray fields that it generates govern the local alignment between the Fe and α–MnAs magnetization directions, which is mostly antiparallel with a marked dependence upon the magnetic domain size.

The complex microstructural and magnetic behavior of metallic MnAs films epitaxially grown on GaAs(001) has been the object of many investigations over the last decade [1]. Indeed, MnAs electrodes, like Fe and FeCo ones, can provide a spin-polarized source for achieving spin injection [2] and tunnel magnetoresistance in GaAs-based spintronic devices [3]. Interestingly, MnAs films display, at room temperature (RT), a self-organized pattern of submicron-wide stripes, alternating the ferromagnetic hexagonal α–MnAs phase with the non-ferromagnetic orthorhombic β–MnAs phase. In general, β–MnAs is considered paramagnetic [4], although a recent study suggests that it may become antiferromagnetic when stripes form [5]. The stripes are aligned parallel to the **c** axis of the α–MnAs phase, whose easy magnetization direction is perpendicular to the stripes. The film thickness (e) controls the period ($T \sim 4.8\ e$) of the self-organized pattern and the height (h) of the steps ($h \sim 0.01\ e$) between ridges and grooves, i.e. the α and β stripes, respectively [1]. For a given thickness, i.e. for a given period, the sample temperature determines the relative width of α and β stripes.

Complex magnetic behaviors have been observed when MnAs is coupled to a ferromagnetic material. Zhu et al., for instance, observed a spin valve effect in self-exchange-biased ferromagnetic MnAs/(Ga,Mn)As FM-metal/FM-semiconductor heterostuctures [6].

Characterized by a regular pattern of sub-micron wide stripes separated by nanometric steps, MnAs/GaAs(001) films can be considered also as self-organized templates for the growth of nanometer-thick magnetic layers, whose magnetic and microstructural properties can be controlled by varying the temperature [7]. Within this framework, we prepared ultra-thin Fe films on MnAs/GaAs(001) templates with the purpose of investigating the influence of the stripe formation and disappearance on the Fe magnetic properties [8]. Taking advantage of the element-selectivity and of the magnetic sensitivity of x-ray resonant magnetic scattering (XRMS), we have already highlighted the complex behavior of the Fe magnetization, showing in particular that the magnetic alignment between Fe and α–MnAs switches from parallel to antiparallel as a function of either the applied field or of the sample temperature. We interpreted our results by assuming that the effective magnetic field felt by the Fe on top of α–MnAs is the sum of the external applied field ($H_{ext}$) and of a dipolar term ($H_D$) generated by the underlying α–MnAs ferromagnetic stripe [8]. Strong applied fields ($H_{ext} > H_D$) impose a parallel alignment between Fe and MnAs, while at lower fields ($H_{ext} < H_D$) the dipolar term dominates, switching the coupling to antiparallel.

In this letter we present an X-ray photoemission electron microscopy (X-PEEM) study of the magnetic coupling between Fe and MnAs/GaAs(001). In addition to element selectivity and magnetic sensitivity, X-PEEM posses the capability of probing magnetic coupling locally, at the level of single domains, representing an ideal complement to our previous spatially averaged XRMS analysis. In particular, X-PEEM studies make it possible to include the details of the magnetic structure in the analysis via micromagnetic calculations.

A 70nm thick MnAs layer was prepared by molecular beam epitaxy on GaAs(001) following the procedure detailed in ref. [8]. The 2 nm thick Fe layer was deposited on top of the MnAs layer kept at 150°C, using a Knudsen cell. Finally, the sample was capped with 1nm of non-reactive ZnSe plus 2nm of Al for protection against oxidation. Transmission electron microscopy shows that the epitaxy relations are $(2\text{-}11)_{Fe}//(1\text{-}100)_{MnAs}//(001)_{GaAs}$ and $[11\text{-}1]_{Fe}//[001]_{MnAs}//[1\text{-}10]_{GaAs}$.

All the X-PEEM measurements were performed at the SIM beamline of the Swiss Light Source (Villigen, Switzerland), at RT and with no applied magnetic field.

X-PEEM images are shown in Fig. 1(a-b), the energy of the circularly polarized x-rays being tuned to the Mn-$L_3$ (a) and to the Fe-$L_3$ (b) resonance. Both images are in fact the difference between two measurements performed with opposite helicity of the incoming photons, a common procedure for highlighting the magnetic structure while minimizing all other sources of contrast. In Fig. 1, the stripes run almost parallel to the vertical direction and photons impinge on the sample at 16° grazing incidence and normal to the stripes, probing the magnetization along the easy axis of α–MnAs. In Fig. 1a, dark and white areas correspond to α–MnAs magnetic domains with opposite orientations of the magnetization, parallel and antiparallel to the photon wave vector. They are separated by gray thin vertical lines, corresponding to non-magnetic β stripes. Comparing the two X-PEEM images, one can notice that there is a clear correlation between α–MnAs domains in Fig. 1a and Fe domains in Fig. 1b, with inverted light/dark patterns. For a better comparison, a rectangular area is outlined and compared on an expanded scale. These images confirm, at the local submicron scale, the dominant RT antiparallel alignment between the magnetization of α–MnAs in the template and of Fe in the overlayer.

Fig. 2a compares line scans obtained along the dashed lines of Fig. 1, at the Mn and Fe resonances. The curves show that the degree of antiparallel alignment is very high, but not quite 100%. For a more quantitative analysis of the images, we define a rate of antiparallel coupling $\Delta$ based on the observation of the Fe/MnAs alignment for every α–MnAs domain, $\Delta = 100\%$ meaning that antiparallel alignment is observed for all the considered domains. The analysis of the entire images of Fig. 1 yields $\Delta = 82\%$. A more detailed information can be obtained by indexing each α–MnAs magnetic domain of Fig. 1a according to its size: the bar diagram of Fig. 2b summarizes the observed rate of antiparallel coupling as a function of the area of the α–MnAs domains. We observe that while the largest domains ($> 0.1$ μm$^2$) display an almost complete antiparallel alignment between α–MnAs and Fe ($\Delta > 96\%$), this is no longer the case when their size shrinks below 0.06 μm$^2$ ($\Delta \approx 70\%$).

As mentioned in the introduction, we intepreted the spatially averaged antiparallel alignment observed by XRMS in terms of the dipolar field generated by α–MnAs acting on the Fe overlayer [8]. In order to check whether this approach is consistent with size dependent data of Fig. 2b, we extended our model to consider a configuration of α–MnAs domains as the one depicted in Fig. 3. Three adjacent α–MnAs domains pertaining to the same stripe, with alternating magnetization direction, have thickness c (=70nm, the layer thickness) and width w (=280nm, the average stripe width at RT). Their size along the stripe direction is **a** for the central domain and **b** for the two external ones. The orientation of the Fe magnetization on top of the central domain will be influenced by the sum of the dipolar fields generated by all three α–MnAs domains, and in particular by its component $B_x$ along the easy magnetization direction. Calculations were performed using the Magnetic Bar Calculator on-line code [9]. Fig. 3 shows how the $B_x$ component evolves in magnitude and sign when the relative size **a/b** of the central and external domains is changed. By taking a constant **b** = 560nm and varying **a** (open circles), one observes that $B_x$ changes sign at around **a** ≈ 200nm. This means that, below this size, the dipolar field generated by the central MnAs domain is overcome by the contributions coming from the neighboring larger domains; therefore, the overall dipolar field will favor a parallel rather than antiparallel alignment of the Fe magnetization with that of its underlying α–MnAs domain. The same can be observed when taking a constant **a**=140nm (filled squares) and varying **b**. The possibility of varying **w** by controlling the sample temperature offers an additional parameter that we do not consider here for discussing our RT data.

Finally, Fig. 4 shows magnetic contrast X-PEEM images taken at RT, after annealing the sample *in situ* at T= 65 °C (complete transition to the β–phase). The α–MnAs magnetic domains are much larger than in Fig. 1, extending across tens of stripes with a width that is often in excess of 1μm. Correspondingly, the degree of Fe/MnAs antiparallel coupling is very high (Δ > 96% over the entire analyzed surface), supporting our interpretation outlined above. One important aspect has not been mentioned yet: our interpretation, as in reference [8], requires the hypothesis of a very weak exchange coupling between the two adjacent Fe and MnAs ferromagnetic layers. This has been verified experimentally to be the case [8], but the origin of this weak exchange coupling remains unclear. All our x-ray absorption and dichroism spectra on previous and present samples are characteristic of metallic materials (see ref. [10] for Mn and ref. [11] for Fe), with no evidence of oxidation. We cannot exclude, though, the formation of a very thin interdiffused interface layer with antiferromagnetic or ferrimagnetic character, such as $Fe_{a-x}Mn_xAs$ (a close to 2, a>x>0) [12] or FeMn [13].

X-PEEM element-selective magnetic microscopy demonstrates the presence of a dominant antiparallel alignment between α–MnAs and Fe at RT, confirming previous spatially averaged XRMS results on Fe/MnAs/GaAs(001). In addition, X-PEEM data show that the degree of antiparallel coupling depends strongly on the size of the magnetic domains, being higher for larger domains. Taking into account the presence of several MnAs domains along a single α–stripe, we have interpreted this result within our previous approach based on the influence on the Fe magnetization direction of MnAs-generated dipolar fields.

Acknolowledgements : We thank F. Nolting for useful discussions. The Agence National de la Recherche (France) and C'Nano IdF provided financial supports via the projects MOMES and TRIPEPS, respectively.

References :

[1] L. Däweritz, Rep. Prog. Phys. 69, 2581 (2006) and references therein. L. Däweritz et al., J. Vac. Sci. Technol. B 23, 1759 (2005).

[2] J.-M. George et al., C.R. Physique 6, 966 (2005) and references therein.

[3] V. Garcia et al., Phys. Rev. Lett. 97, 246802 (2006).

[4] C. P. Bean and D. S. Rodbell, Phys. Rev. 126, 104 (1962).

[5] E. Bauer et al., J. Vac. Sci. Technol. B 25, 1470 (2007)

[6] M. Zhu et al., Appl. Phys. Lett. 91, 192503 (2007)

[7] L. N. Coelho et al., J. Appl. Phys. 100, 083906 (2006).

[8] M. Sacchi et al., Phys. Rev. B 77, 165317 (2008).

[9] Magnetic Bar Calculator, Integrated Engineering Software (see http://www.integratedsoft.com/papers/Benchmark/Bar_Magnet.pdf and http://www.integratedsoft.com/calculator.htm )

[10] V. Garcia et al., Phys. Rev. Lett. 99, 117205, (2007).

[11] M. Marangolo et al., Phys. Rev. Lett. 88, 217202 (2002)

[12] T. Kanomata, T. Goto and H. Ido J. Phys. Soc. of Japan, 43 1178 (1977)

[13] F. Offi et al., Phys. Rev. B 67, 094419 (2003)

FIGURE CAPTIONS

Figure 1. Magnetic contrast X-PEEM images obtained by tuning the photon energy at the Mn (a) and Fe (b) $L_3$ resonance. The areas delimited by rectangles are shown on an enlarged scale. White dashed lines correspond to data shown in Fig. 2a.

Figure 2. (a): intensity profiles taken along the dashed lines shown in the X-PEEM images of Fig. 1, at the Mn and Fe 2p resonances. (b): Percentage of antiparallel aligned Fe and α–MnAs magnetic domains as a function of the domain area. Data obtained from images shown in Fig. 1.

Figure 3. Bx is the component perpendicular to the stripes of the dipolar field generated at the position of the central Fe layer by three α–MnAs domains (see sketch on top of figure). The width of the α–MnAs stripe is **w** = 280 nm and its thickness is **c** = 70 nm. Bx is calculated as a function of **a/b**, either for a constant **a** = 140 nm (squares) or for a constant **b** = 560 nm.

Figure 4. Same as Figure 1, after an *in situ* annealing of the sample at T = 65 °C.

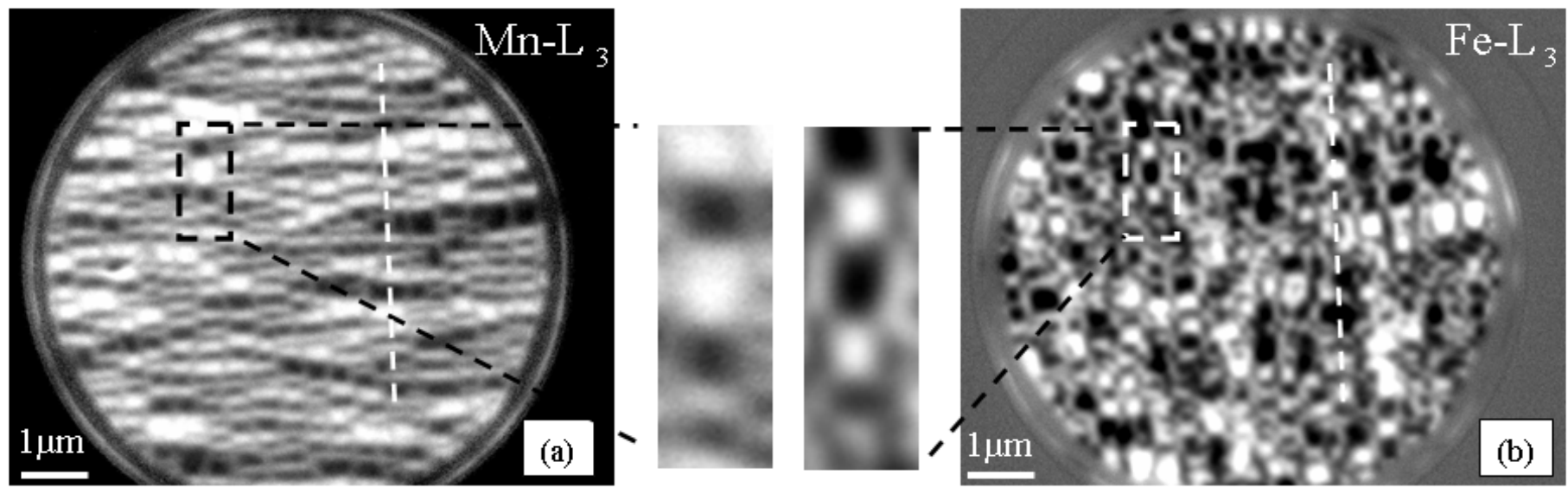

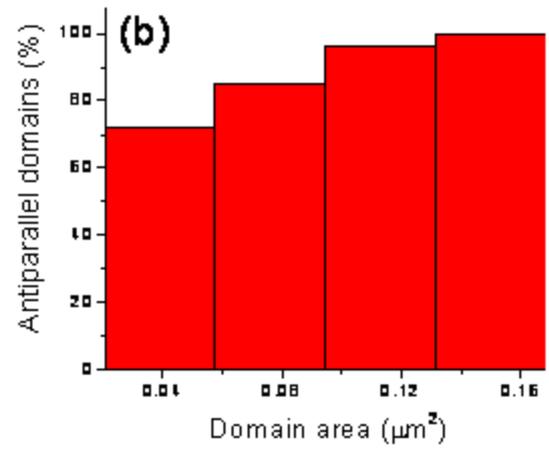

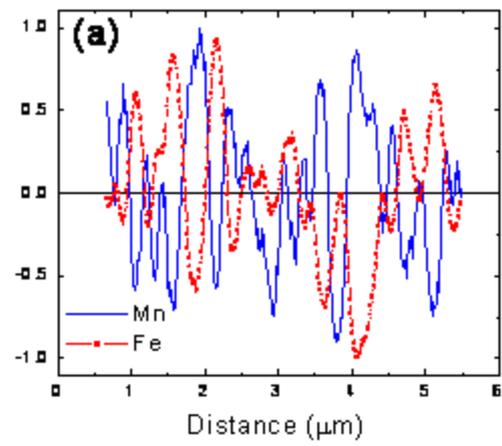

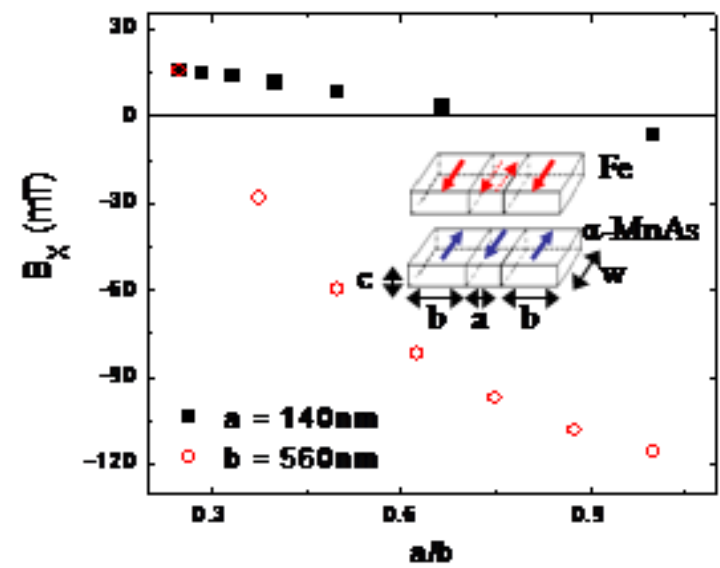

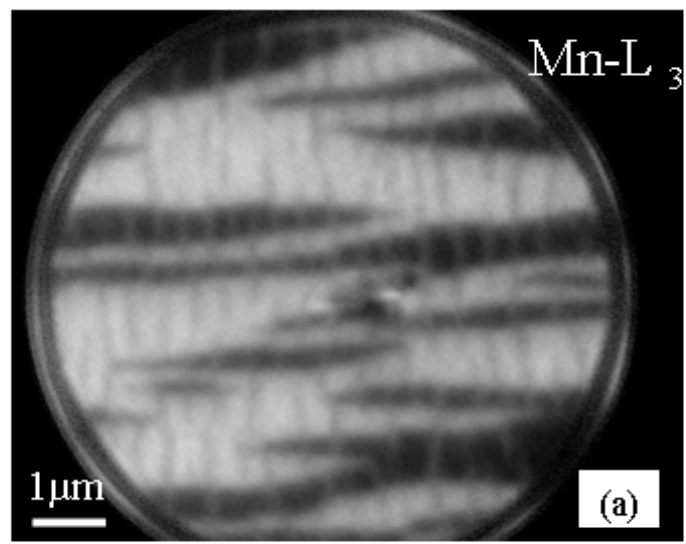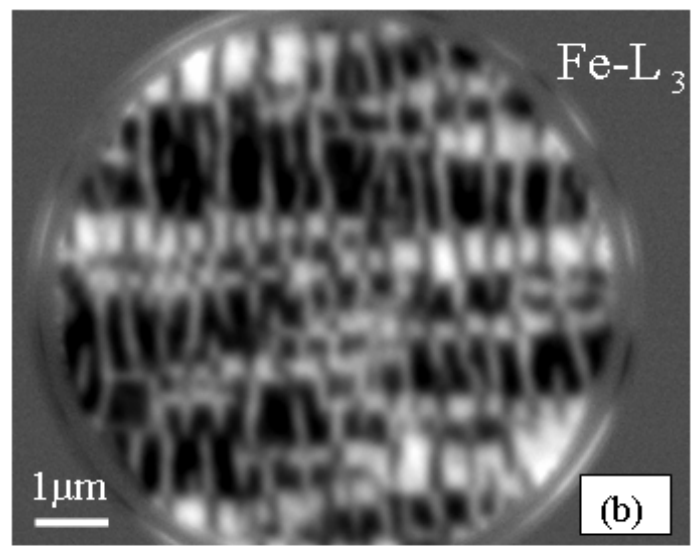